\documentclass[prl,10pt,twocolumn,superscriptaddress,aps,floatfix]{revtex4-1}
\usepackage{graphicx}
\usepackage{natbib}
\usepackage{url}
\usepackage[normalem]{ulem}
\usepackage{color}
\usepackage{amsmath}

\begin{document}

\title{Universality of Electron Mobility in LaAlO$_3$/SrTiO$_3$ and bulk SrTiO$_3$}
\author{Felix Trier}%
\email{Email: felix.trier@cnrs-thales.fr}
\altaffiliation{%
  Current address: Unit\'e Mixte de Physique CNRS, Thales, University Paris-Sud, Universit\'e Paris-Saclay, 91767 Palaiseau, France 
}%
\affiliation{%
  Department of Energy Conversion and Storage, Technical University of Denmark, Ris{\o} Campus, 4000 Roskilde, Denmark.
}%
\author{K. V. Reich}%
\affiliation{%
  Fine Theoretical Physics Institute, University of Minnesota, Minneapolis, Minnesota 55455, USA
}%
\affiliation{%
  Ioffe Institute, St Petersburg, 194021, Russia
}%

\author{Dennis Valbj{\o}rn Christensen}%
\affiliation{%
  Department of Energy Conversion and Storage, Technical University of Denmark, Ris{\o} Campus, 4000 Roskilde, Denmark.
}%
\author{Yu Zhang}%
\affiliation{%
  Department of Energy Conversion and Storage, Technical University of Denmark, Ris{\o} Campus, 4000 Roskilde, Denmark.
}%
\author{Harry L. Tuller}%
\affiliation{%
  Department of Materials Science and Engineering, Massachusetts Institute of Technology, Cambridge, Massachusetts 02139, USA
}%
\author{Yunzhong Chen}%
\affiliation{%
  Department of Energy Conversion and Storage, Technical University of Denmark, Ris{\o} Campus, 4000 Roskilde, Denmark.
}%
\author{B. I. Shklovskii}
\affiliation{%
  Fine Theoretical Physics Institute, University of Minnesota, Minneapolis, Minnesota 55455, USA
}%
\author{Nini Pryds}%
\email{Email: nipr@dtu.dk}
\affiliation{%
  Department of Energy Conversion and Storage, Technical University of Denmark, Ris{\o} Campus, 4000 Roskilde, Denmark.
}%
\date{\today}

\begin{abstract}
  Metallic LaAlO$_3$/SrTiO$_3$ (LAO/STO) interfaces attract enormous attention, but the relationship between the electron mobility and the sheet electron density, $n_s$, is poorly understood. Here we derive a simple expression for the three-dimensional electron density near the interface, $n_{3D}$, as a function of $n_s$ and find that the mobility for LAO/STO-based interfaces depends on $n_{3D}$ in the same way as it does for bulk doped STO. It is known that undoped bulk STO is strongly compensated with $N \simeq  5 \times 10^{18}~\rm{cm^{-3}}$ background donors and acceptors. In intentionally doped bulk STO with a concentration of electrons $n_{3D} < N$ background impurities determine the electron scattering. Thus, when $n_{3D} < N$ it is natural to see in LAO/STO the same mobility as in the bulk. On the other hand, in the bulk samples with $n_{3D} > N$ the mobility collapses because scattering happens on $n_{3D}$ intentionally introduced donors. For LAO/STO the polar catastrophe which provides electrons is not supposed to provide equal number of random donors and thus the mobility should be larger. The fact that the mobility is still the same implies that for the LAO/STO the polar catastrophe model should be revisited.
\end{abstract}

\maketitle

Recently, much attention has been directed at metallic LaAlO$_3$/SrTiO$_3$ (LAO/STO) interfaces\cite{mannhart_oxide_2010, schlom_oxide_2011, sulpizio_nanoscale_2014, gabay_oxide_2013, cheng_electron_2015}. Such interfaces have been shown to exhibit a plethora of physical phenomena. These include gate-tunable superconductivity, magnetism, metal-insulator transitions and quantized Hall resistance \cite{reyren_superconducting_2007, caviglia_electric_2008, prawiroatmodjo_evidence_2016, brinkman_magnetic_2007, thiel_tunable_2006, christensen_controlling_2013, trier_quantization_2016, matsubara_observation_2016}. Motivated by interest in quantum phenomena and potential applications, multiple studies have specifically sought to improve the low-temperature electron mobility in LAO/STO interfaces \cite{chen_high-mobility_2013, chen_creation_2015, chen_extreme_2015, kozuka_enhancing_2010, matsubara_observation_2016, huijben_defect_2013, xie_enhancing_2013, fete_growth-induced_2015, zeng_liquid-gated_2016}. In spite of more than a decade of research, the dominant scattering mechanism in LAO/STO interfaces, nonetheless, remains elusive.

\begin{figure}
  \includegraphics[width=0.48\textwidth]{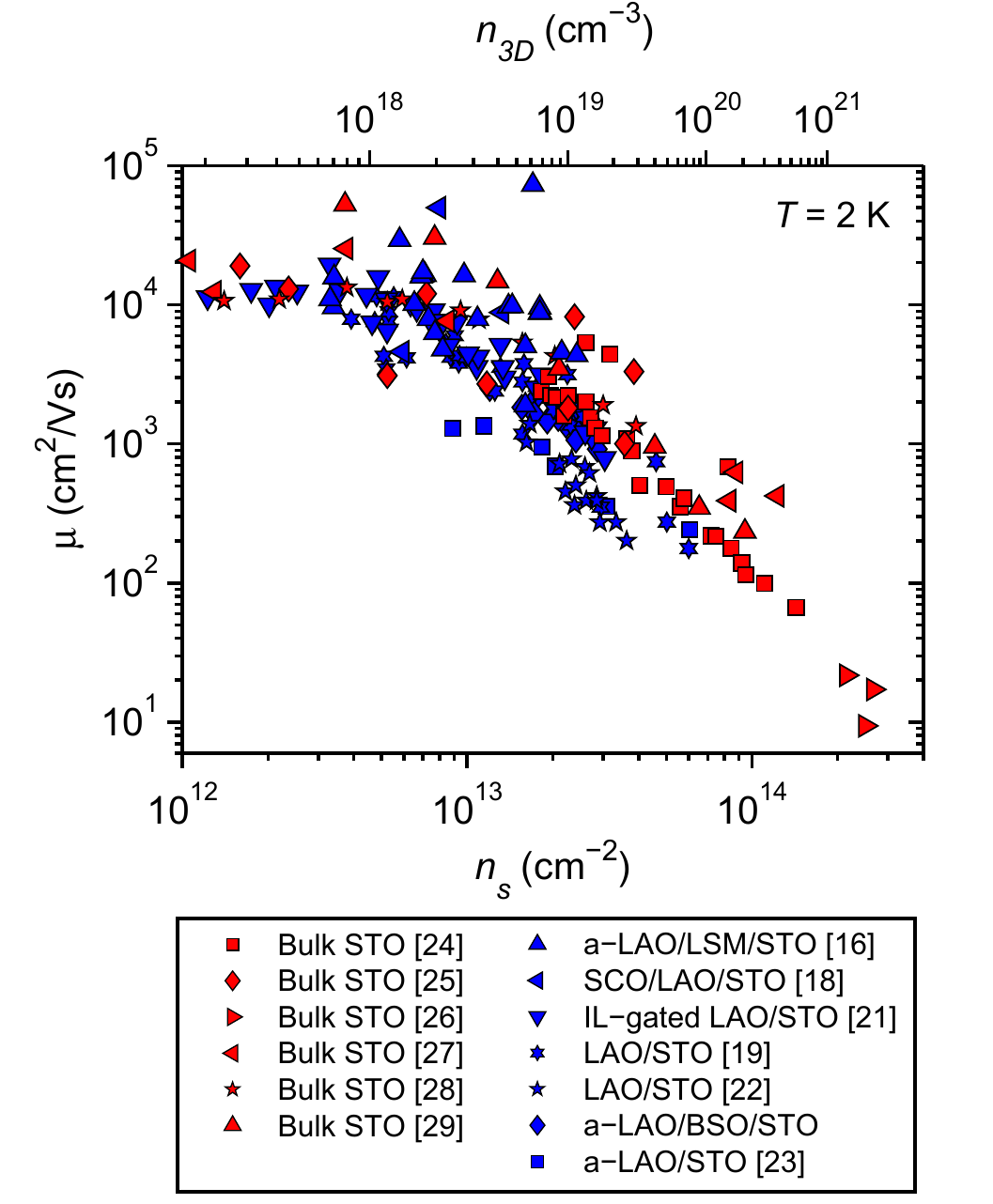}
  \caption{(Color online) The dependence of the low-temperature electron mobility, $\mu$, on the sheet carrier density, $n_s$, (lower scale) and  bulk carrier density, $n_{3D}$, (upper scale) . The data for bulk STO (red markers) and LAO/STO-based interfaces (blue markers) are close to each other. The  data encompass the following STO-based systems: bulk STO \cite{koonce_superconducting_1967, tufte_electron_1967, gong_oxygen-deficient_1991, spinelli_electronic_2010, verma_intrinsic_2014, lin_metallicity_2017}; amorphous-LaAlO$_3$/LaSrMnO$_{3}$/SrTiO$_{3}$ (a-LAO/LSM/STO) \cite{chen_extreme_2015} and this work; SrCuO$_2$/LaAlO$_3$/SrTiO$_{3}$ (SCO/LAO/STO) \cite{huijben_defect_2013}; ion liquid-gated LaAlO$_3$/SrTiO$_{3}$ (IL-gated LAO/STO) \cite{zeng_liquid-gated_2016}; LaAlO$_3$/SrTiO$_{3}$ (LAO/STO) \cite{xie_enhancing_2013, sanders_laalo_2015}; amorphous-LaAlO$_3$/BaSnMnO$_{3}$/SrTiO$_{3}$ (a-LAO/BSO/STO) this work; amorphous-LaAlO$_3$/SrTiO$_{3}$ (a-LAO/STO) \cite{chen_metallic_2011} and this work. }.
  \label{figure1}
\end{figure}

The canonical LaAlO$_3$/SrTiO$_3$ interface (LAO/STO) has previously been shown to exhibit large variations of the low-temperature electron mobility $\mu$ and sheet electron density $n_s$, which depend strongly on interface growth conditions \cite{xie_enhancing_2013, sanders_laalo_2015}. Remarkably in these previous studies, the dependence of $\mu(n_s)$ seems to follow a universal behaviour \cite{xie_enhancing_2013, sanders_laalo_2015}, almost regardless of the type of LAO/STO interface and particular growth conditions (see blue markers in Fig. \ref{figure1}).

The aim of this paper is to relate $\mu$($n_s$) data for LAO/STO-based interfaces from the literature \cite{chen_extreme_2015, huijben_defect_2013, kozuka_enhancing_2010, matsubara_observation_2016, xie_enhancing_2013, sanders_laalo_2015, zeng_liquid-gated_2016, chen_metallic_2011} and from newly prepared by us LAO/STO-based samples with the available electron mobility data for bulk doped STO over a range of three-dimensional electron densities from the literature \cite{koonce_superconducting_1967, tufte_electron_1967, gong_oxygen-deficient_1991, spinelli_electronic_2010, lin_metallicity_2017, verma_intrinsic_2014} (see red markers in Fig. \ref{figure1}). To this end, we note that at low temperature the electrons are distributed in a layer of width, $d \simeq 5-100$ nm \cite{Hwang_Xray, Hwang_PL, LAO_STO_Berreman, abinitio_STO, abinitio_STO_2} near the LAO/STO interface. Due to the relatively large width of this electron layer we essentially deal with a three-dimensional system \cite{MacDonald_theory,distribution_LAO_STO}. Such an electronic system can be described by the recent theory \cite{reich_accumulation_2015, Collapse_Han, Han_review} of accumulation layers in STO based on a combination of the Landau-Ginzburg description of the dielectric response of the STO lattice and the Thomas-Fermi approximation for the degenerate electron gas. Using this theory we show below that the three-dimensional electron density of the electronic system near the LAO/STO interface, $n_{3D}$, depends on the measured two-dimensional sheet electron density, $n_s$, in the following way:
\begin{equation}
      \label{eq:3Dconcetration}
      n_{3D}(n_s)= \frac{C (n_sa^{2})^{6/5}}{(a_{B}a^{4})^{3/5}} \left[1+\frac{A \kappa }{8\pi}(n_sa^{2})^{2}\right]^{3/5}.
    \end{equation}
    Here $C \simeq (5^{3} 2^{3} 3^{-2} \pi^{-1})^{1/5} \simeq 2$, $a \simeq 3.9$ \AA\ is the lattice constant of STO, $a_{B}=\hbar^{2} 4\pi \varepsilon_{0} \kappa/m^\star e^{2} \simeq 7000$ \AA\ is the effective Bohr radius, $\kappa \simeq 20000$ is the dielectric constant of STO at low temperatures, the coefficient $A \simeq 0.9$ describes the non-linear dielectric response of STO and $m^\star \simeq 1.6m_{e}$ is the effective electron mass in STO \cite{caviglia_two-dimensional_2010, ben_shalom_shubnikovhaas_2010, trier_quantization_2016}, with $m_{e}$ being the free electron mass. For example, a three-dimensional density of $n_{3D} = 5\times 10^{18}$ cm$^{-3}$ corresponds to a two-dimensional sheet density of $n_s=1.5\times 10^{13}$ cm$^{-2}$.

In Fig. \ref{figure1} by using Eq. \eqref{eq:3Dconcetration} we compare the low temperature electron mobility for LAO/STO-based interfaces with mobility for bulk doped STO having the same electron density $n_{3D}$. We see that the values of these mobilities and the concentration dependencies  for LAO/STO-based interfaces and bulk STO are similar. This  suggests that the electron scattering mechanisms in LAO/STO-based interfaces are the same as those in bulk STO.

In the following we briefly review known mechanisms for electron scattering in bulk conducting STO at low temperatures \cite{tufte_electron_1967,verma_intrinsic_2014}. Let us first dwell on the interpretation of the electron mobility in nominally insulating bulk STO. It is well established that bulk STO samples usually are strongly compensated \cite{tufte_electron_1967, spinelli_electronic_2010, rice_persistent_2014}, i.e. have nearly equal donor and acceptor concentrations. In bulk STO, the possible background donors include oxygen vacancies, while common acceptors include strontium vacancies, aluminium or iron \cite{xu_disentanglement_2016, rice_persistent_2014,Ambwani_2016}. Due to a transition of electrons from donors to acceptors, the total concentration of charged background impurities can be as large as $N \sim 5 \times 10^{18}$ cm$^{-3}$ \cite{rice_persistent_2014}. This means that both in bulk STO and LAO/STO-based interfaces  with $n_{3D}\ll N$ (or $n_s \ll 1.5 \times 10^{13}$ cm$^{-3}$) the electron mobility is limited by electrons scattering on background charged impurities. For bulk STO having $n_{3D} \gg N$  (or $n_s \gg 1.5 \times 10^{13}$ cm$^{-3}$) , the low-temperature electron mobility is instead limited by scattering on intentionally added ionized donors. On one hand these ionized donors provide free charge carriers to the electronic system, on the other hand, they act as scattering centres. If we assume that in LAO/STO-based interfaces with $n_{3D} > N$ electrons are provided by a conventional polar catastrophe, which does not bring positive random donors together with them;  mobility of such samples should be much larger than for the bulk STO with the same 3D concentration. Thus, the overall universality of the electron mobility in Fig. \ref{figure1} allows us to conclude that electrons in LAO/STO-based interfaces with $n_{3D} > N$ are provided not by the polarization catastrophe, but by the equal number of donors which scatter electrons. This observation questions the polarization catastrophe model and requires its further adjustment. For the case of amorphous-LAO/STO grown at room temperature, the role of  ionized donors is undoubtedly played by oxygen vacancies in STO located near the interface \cite{chen_metallic_2011}.

Let us now dwell on our experimental details. The a-LAO/STO, a-LAO/BSO/STO and a-LAO/LSM/STO samples included in this study were all prepared on TiO$_2$-terminated and (100)-oriented SrTiO$_3$ (STO) substrates (5$\times$5$\times$0.5 mm$^3$, miscut angle $<$ 0.2$^\circ$). For the a-LAO/BSO/STO and a-LAO/LSM/STO samples, a single unit cell epitaxial BaSnO$_3$ (BSO) or La$_{7/8}$Sr$_{1/8}$MnO$_3$ (LSM) spacer layer, respectively, was deposited by pulsed laser deposition (PLD) in an oxygen atmosphere of $\sim$10$^{-4}$ mbar at 600 $^\circ$C. The film growth rate of LSM and BSO was determined from \textit{in-situ} RHEED oscillations. Furthermore, the LSM and BSO were ablated from ceramic LSM and BSO targets with a target-substrate distance of 5.7 cm, by a KrF laser ($\lambda$ = 248 nm) with repetition rate of 1 Hz and laser fluence of 1.5 Jcm$^{-2}$. The respective samples were then cooled under $\sim$10$^{-4}$ mbar oxygen pressure at a rate of 15 $^\circ$C/min to room temperature ($<$25 $^\circ$C). Finally, 16 nm amorphous-LaAlO$_3$ (a-LAO) was grown at an oxygen pressure of $\sim$10$^{-6}$ mbar and room temperature on the respective samples to finalize the a-LAO/STO, a-LAO/BSO/STO and a-LAO/LSM/STO heterostructures. The a-LAO was ablated from a single crystalline LAO target and otherwise identical PLD conditions were used as for BSO and LSM. All samples were mounted on ceramic chip-carriers and electrically connected in the van der Pauw geometry. The interface of all samples was contacted using ultrasonically wire-bonded aluminum wires. Sheet resistance and Hall resistance measurements between 2--300 K were performed using a standard DC technique ($I_{DC}$ = 1--5 $\mu$A) in a cryostat with magnetic fields up to 15 T.

Note that all the used in Fig. \ref{figure1} data are obtained for the LAO layers grown on the (100) oriented STO. Although thickness of these LAO layers varies, the mobility has the same dependence on concentration.We did not include LAO layers grown on (110) or (111) oriented STO, such as Ref. \onlinecite{Herranz_2012}. We found that even in such samples the dependence of the mobility on the concentration is the same as in Fig. \ref{figure1}, with exception of the thickest sample. It shows very small mobility, possibly because of large surface roughness.

Below, we present the derivation of Eq. \eqref{eq:3Dconcetration}, which allowed LAO/STO versus bulk STO mobility comparison. The three-dimensional distribution of electrons near the STO interface, $n(x)$, was found in Ref. \cite{reich_accumulation_2015} in the two limiting cases of small and large two-dimensional electron densities $n_s$. For our purpose, we need the three-dimensional electron density near the interface  $n_{3D}=n(0)$ for any $n_s$. Thus we have to return to this problem again.

We are interested in the accumulation layer near an interface of STO. We then consider the case when the axis $x$ is directed perpendicular to the interface (plane $x=0$) and lies along the [100] axis of a cubic crystal of STO. Electrons are located near the surface due to the attractive potential of positive charges $en_s$ near the interface. These charges create an external field $D_0 = 4\pi e n_s$ (here and below we use cgs units) applied from the outside of STO, which is directed along the $x$ axis. In that case, the problem is effectively one-dimensional. If the electron three-dimensional density is denoted by $n(x)$, then the potential depth profile $\varphi(x)$ in the system is determined by the equations:
\begin{eqnarray}
  \label{eq:Gauss}
  \frac{d D}{dx}   = - 4 \pi e  n(x) ,  ~  D =E+4\pi P , ~   \frac {d \varphi}{d x}  = -E ,
\end{eqnarray}
where $D(x), E(x), P(x)$ are the electric displacement, the electric field and the polarization in STO, respectively. Eq. \eqref{eq:Gauss} should be solved using proper boundary conditions. For an accumulation layer the boundary conditions are $D(0)=D_0$ and $\varphi(\infty)=0$.

To solve the system of equations \eqref{eq:Gauss} one also needs to know the two material relationships $E(P)$ and $\rho(\varphi)$. Let us start from the lattice dielectric response $E(P)$. STO is well known as a quantum paraelectric, where the onset of ferroelectric order is suppressed by quantum fluctuations. A powerful approach to describe the properties of ferroelectric-like materials is based on the Landau-Ginzburg theory. For a continuous second-order phase transition the Landau-Ginzburg expression of the free energy density $F(x)$ is represented as a power series expansion with respect to the polarization $P$:
\begin{equation}
  \label{eq:F}
  F(x)= F_0 + \frac{2\pi}{\kappa} P(x)^2 + \frac{1}{4} A \frac{1}{P_0^2} P(x)^4,
\end{equation}
where $F_0$ stands for the free energy density at $P = 0$, $P_0=e/a^2$ is the characteristic polarization and the coefficient A describes the non-linear dielectric response. In general $F$ depends on the components of the vector $P$, but in the chosen geometry the problem is one-dimensional, and all vectors are directed along the $x$ axis. The crystal polarization $P$ is determined by minimizing the free energy density $F$ in the presence of the electric field $E$ where $\delta F/ \delta P= E$. This condition relates $E$ and $P$. We note that $E \ll 4\pi P$ and thus $D=E+4\pi P \simeq 4\pi P$.
Due to electric neutrality, the number of accumulated electrons has to compensate the external field $D_0$, i.e., 
\begin{equation}
  \label{eq:neutrality}
  D_0 = 4 \pi e n_s = 4 \pi e \int \limits_0^{\infty} n(x) dx.
\end{equation}
To take into account the electron screening of the external field we use the Thomas-Fermi approach in which the electron concentration $n(x)$ and the self-consistent potential profile $\varphi(x)$ are related as $e\varphi(x)+\varepsilon(x)=\varepsilon_F=0$, where 
\begin{equation}
  \label{eq:TF}
  \varepsilon(x)= (3\pi^2)^{2/3}\frac{\hbar^2}{2 m^\star} [n(x)]^{2/3}
\end{equation}
is the chemical potential of the electron gas. (Here we consider lightly doped STO) Using Eqs \eqref{eq:F}, \eqref{eq:TF} and \eqref{eq:Gauss} we arrive at:

\begin{equation}
  \label{eq:n_F}
  F(x)-F_{0} = \frac{3^{2/3}\pi^{4/3}}{5} \frac{\hbar^{2}}{m^\star} n(x)^{5/3}.
\end{equation}
Using this relationship at the surface $x=0$ with  $P(0) \equiv en_s$ and $n(0) \equiv n_{3D}$ we arrive at Eq. \eqref{eq:3Dconcetration}.

To conclude, we find that the dependence of the electron mobility, $\mu$, for LAO/STO-based interfaces on the three-dimensional electron density near the interface, $n_{3D}$, is akin to the electron mobility dependence on density for bulk doped STO. This observation implies that the same scattering mechanism dominates the charge transport in both systems. In particular, at low electron densities $n_{3D} < N$ and the electron mobility is limited by background impurity concentrations in STO. On the other hand in the bulk samples with $n_{3D} > N$ the mobility collapses because scattering happens on intentionally introduced donors. For LAO/STO the polar catastrophe which provides electrons is not supposed to provide random donors. The fact that nevertheless the mobility is the same says that for LAO/STO the model of disorder free  polar catastrophe should be revised.

\begin{acknowledgments}
  The theoretical work was supported by the National Science Foundation through the University of Minnesota MRSEC under Award Number DMR-1420013. H. Tuller thanks the National Science Foundation for his support under Award number DMR-1507047. K.V.R was supported by the Russian Science Foundation under grant 17-72-10072. The authors gratefully acknowledge discussions with M. Bibes, C. Leighton, B. Jalan, H. Fu and D.C. Vaz.
\end{acknowledgments}

\end{document}